\documentclass[%
reprint,
superscriptaddress,
amsmath,amssymb,
aps,
pre
]{revtex4-1}

\usepackage{graphicx}
\usepackage{comment}
\usepackage{dcolumn}
\usepackage{bm}
\usepackage{color}

\begin{document}

\preprint{APS/123-QED}

\title{Spontaneous deformation and fission of oil droplets on an aqueous surfactant solution}

\author{Masahide Okada}%
\affiliation{%
 Department of Physics, Chiba University, Chiba 263-8522, Japan
}%

\author{Yutaka Sumino}
\affiliation{%
 Department of Applied Physics, Tokyo University of Science, Tokyo 125-8585, Japan
}%
\affiliation{
 W-FST, I$^2$ Plus, and DCIS, RIST, Tokyo University of Science, Tokyo 125-8585, Japan
}
\author{Hiroaki Ito}%
\affiliation{%
 Department of Physics, Chiba University, Chiba 263-8522, Japan
}%

\author{Hiroyuki Kitahata}
\email{kitahata@chiba-u.jp}
\affiliation{%
 Department of Physics, Chiba University, Chiba 263-8522, Japan
}%

\date{\today}

\begin{abstract}
 We investigated the spontaneous deformation and fission of a tetradecane droplet containing palmitic acid (PA) on a stearyltrimethylammonium chloride (STAC) aqueous solution.
 In this system, the generation and rupture of the gel layer composed of PA and STAC induce the droplet deformation and fission.
 To investigate the characteristics of the droplet-fission dynamics, we obtained the time series of the number of the droplets produced by fission and confirmed that the number has a peak at a certain STAC concentration.
 Since the fission of the droplet should be led by the deformation, we analyzed four parameters which may relate to the fission dynamics from the spatiotemporal correlation of the droplet-boundary velocity.
 We found that the parameter which corresponds to the expansion speed had the strongest positive correlation among them, and thus we concluded that the faster deformation would be the key factor for the fission dynamics.
 
\end{abstract}

\pacs{Valid PACS appear here}
\maketitle



\section{Introduction}
\label{Intro}
 Self-propelled objects have been intensively studied as examples of nonequilibrium systems.
 These objects move through the transduction of chemical energy to mechanical energy.
 For example, alcohol droplets\,\cite{pre05,jcp16,Nat15} and camphor disks\,\cite{lang97,lang01} on an aqueous phase, alcohol droplets on an oil phase\,\cite{prl17}, soap disks at  an oil-water interface\,\cite{cpl05,jcp11}, aqueous droplets with a periodic chemical reaction\,\cite{cl12,pccp14}, and oil droplets in a surfactant aqueous solution\,\cite{acie11,cpc13,jacs07} are typical self-propelled objects.
 They are driven by the surface-tension gradient originating from the concentration gradient of the chemicals.
 In the case that the driving force overcomes the interfacial tension that tends to keep the droplet shape spherical, they exhibit shape deformation as well as motion.

 Through dynamic deformation, the liquid droplets can also exhibit fission\,\cite{pre05,jcp16,prl17,cl12,pccp14,acie11,cpc13}.
 Nagai et al. reported the spontaneous motion, deformation, and fission of a pentanol droplet on an aqueous surface, and performed the stability analyses\,\cite{pre05}.
 Keiser et al. discussed the spread and successive fragmentation at the periphery of an alcohol-aqueous-solution droplet on an oil phase based on the experimental results and scaling analyses\,\cite{prl17}.
 In these systems, once the periphery of the alcohol droplet is rippled, such deformation grows due to the instability induced by the interfacial-tension gradient and finally leads to the droplet fission.
 Many other studies have also reported the droplet fission due to the Marangoni effect\,\cite{jcp16,cl12,pccp14,acie11,cpc13,plb83}.
 
 It was reported that a tetradecane droplet containing palmitic acid (PA) on an aqueous solution of stearyltrimethylammonium chloride (STAC) exhibits spontaneous blebbing and fission\,\cite{pre07,rscc19}. 
 Here, blebbing means the formation of a bleb, a spherical deformation that resembles ones observed in cell deformation\,\cite{jm08}.
 It is suggested that the blebbing observed in the PA-STAC system is caused by the formation of the gel layer with lamellar structures composed of PA and STAC, which was confirmed by time-dependent small-angle x-ray scattering (SAXS)\,\cite{lang12,lang16}.
 
 A bleb expands by the increase of internal pressure caused by the formation of the gel layer.
 It shrinks by the increase of effective interfacial tension caused by the rupture of the gel layer.
 The mechanisms for the expansion and the shrinkage are not symmetric, and the droplet does not recover its shape after the blebbing.
 As a result, we observe the significant distortion of the droplet after the repetition of the blebbing.
 Furthermore, the large distortion may lead spontaneous fission of the droplet\,\cite{pre07,jpcb09}.
 Note that the typical timescale of the oil droplet fission led by the blebbing $\sim 10^1$--$10^3\,\mathrm{s}$ is longer than that led by the Marangoni effect $\sim 10^0$--$10^1\,\mathrm{s}$ in the systems with the lengthscale of mm.
 Detailed analyses of the droplet deformation focusing on the surface velocity were performed and it was confirmed that the droplet exhibits blebbing irregularly, that is to say, neither long-range spatial correlation nor long-time temporal correlation was observed\,\cite{soft11}.
 The surfactant-concentration dependence of the droplet blebbing was also studied, and the droplet behaviors were classified into the three types depending on the concentrations: fission induced by interfacial blebbing, interfacial blebbing without fission, and the formation of the white-turbid aggregate that covers the droplet\,\cite{jpcb09}.
 
 Every time we observed the fission of the oil droplet, it exhibited active and frequent blebbing before the fission.
 Thus, we assumed that the fission of the droplet was related with the droplet deformation.
 In the previous studies, the droplet deformation was focused on.
 In the present study, we carried out the detailed analysis of the oil droplet fission caused by the generation and rupture of the gel layer in the same system as the previous studies\,\cite{pre07,jpcb09,soft11,lang12,lang16}, and discussed the droplet fission depending on the STAC concentration.
 We experimentally obtained the time series of the number of the droplets for various STAC concentrations.
 To discuss the fission dynamics, we investigated the dynamics of the droplet deformation for various STAC concentrations and extracted four characteristic parameters which seem to relate with the mechanism of fission from the droplet-deformation dynamics.
 From these investigations, we revealed the relationship between deformation and fission, and found the key parameter that may affect the droplet fission.
 

\section{Experimental setup}
\label{setup}
\begin{figure}[!ht]
\center
\includegraphics{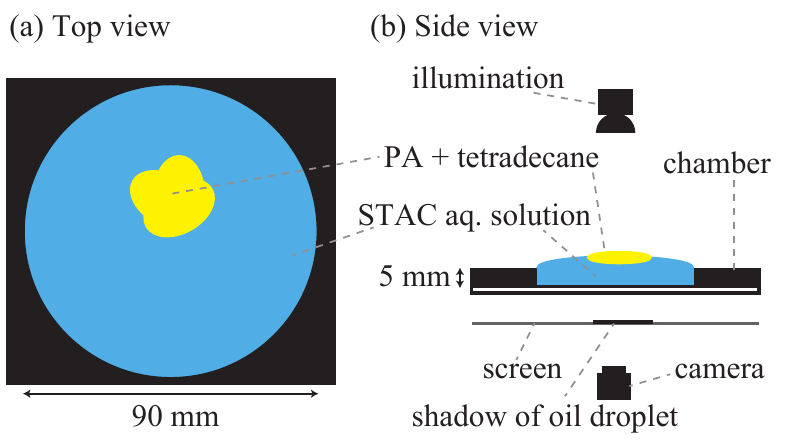}
\caption{Experimental setup. (a)~Top view. (b)~Side view. The chamber was made of a teflon plate with a circular hole (diameter: 90 mm, thickness: 5 mm) attached to an acrylic plate. The chamber was filled with STAC aqueous solution with a volume of $38\,\mathrm{ml}$. An oil droplet with a volume of $0.8\,\mathrm{ml}$ was put on the surface of the STAC solution.}
\label{st} 
\end{figure}

 STAC and PA were purchased from Tokyo Chemical Industry Co., Ltd., Tokyo, Japan.
 Tetradecane was purchased from Sigma-Aldrich, St. Louis, USA.
 Water was purified by Elix UV 3 (Merck, Darmstadt, Germany).
 We prepared a teflon plate with a circular hole (diameter:$\, 90\, \mathrm{mm}$, thickness:$\,5\, \mathrm{mm}$) attached to an acrylic plate, which was used as a chamber.
 Experimental setup is schematically illustrated in Fig.~\ref{st}.
 The chamber was filled with $38\,\mathrm{ml}$ of STAC aqueous solution, and $0.8\,\mathrm{ml}$ of PA tetradecane solution was put as a droplet on the aqueous surface. 
 The concentration of STAC, $C$, was varied from $0.2$ to $5\,\mathrm{mM}$, while the concentration of PA was fixed at $20\,\mathrm{mM}$.
 In this condition, it was reported that the droplets deform and exhibit fission\,\cite{jpcb09}.
 
 We put a tracing paper as a screen under the chamber and illuminated the oil droplet from the top.
 We recorded the shadow of the droplet on the screen from the bottom using a USB camera (DMK 24UJ003; Imaging Source Asia Co., Ltd., Taipei, Taiwan) at 1 fps.
 All experiments were carried out at room temperature ($23.5 \pm 1.5\,^\circ \mathrm{C}$).

\section{Experimental results}
\label{result}
 
\begin{figure*}[bt]
\center
\includegraphics{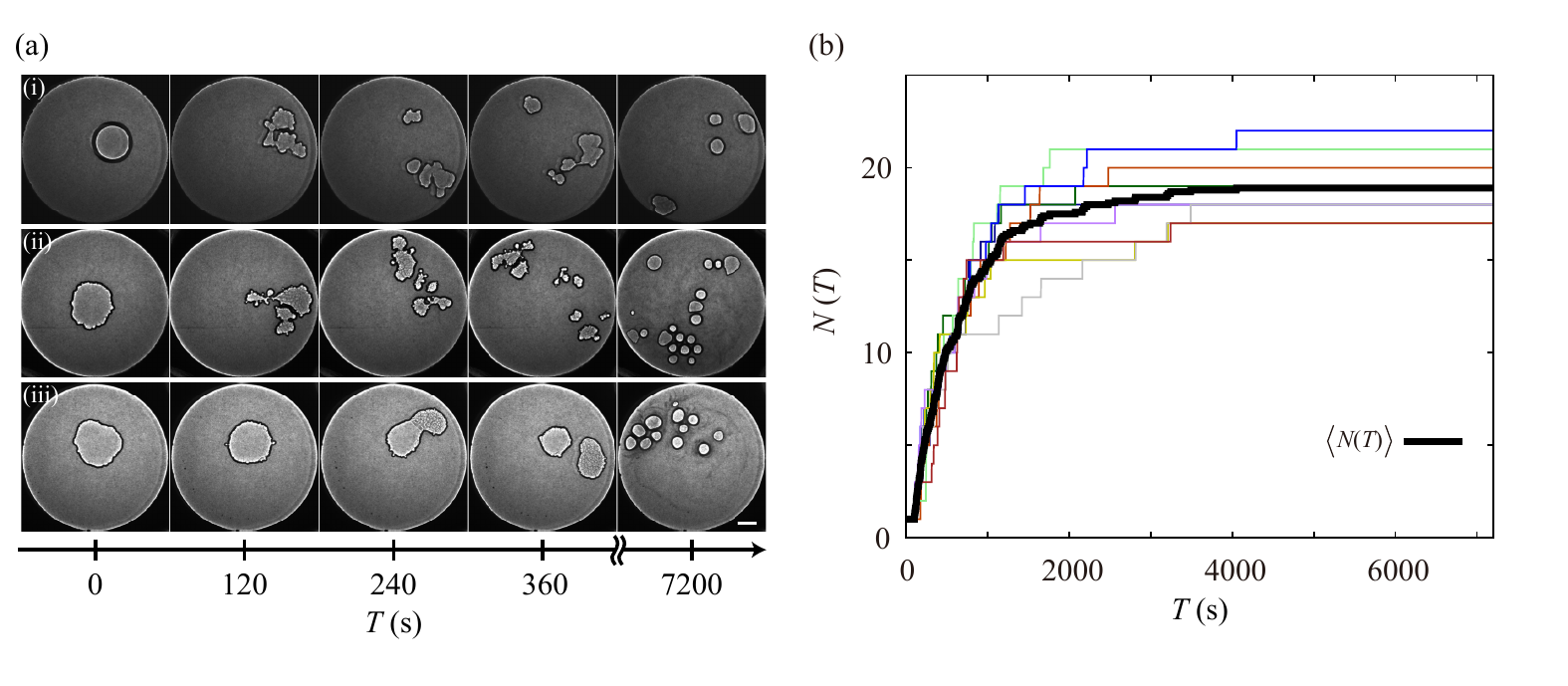}
\caption{Experimental results on the droplet fission. (a)~Snapshots for various $C$. Scale bar: $10\, \mathrm{mm}$. $C=$ (i)\,$0.2\,\mathrm{mM}$, (ii)\,$2\,\mathrm{mM}$, (iii)\,$5\,\mathrm{mM}$. (b)~Time series of the number of the oil droplets for $C=2\,\mathrm{mM}$. The thin lines show the experimental results for ten trials, and the thick line shows the mean value.}
\label{snap} 
\end{figure*}
 
 After putting the oil droplet on the aqueous phase, it had a circular shape during an induction period ($\sim10\,\mathrm{s}$)\,\cite{pre07}.
 After the induction period, the area of the oil droplet quickly expanded and then shrank within a few seconds\,\cite{jpcb09}.
 After these behaviors, blebbing started, where the droplet boundaries exhibited slow circular expansion followed by rapid shrinkage in a spatiotemporally random manner\,\cite{pre07}.
 We defined $T=0\,\mathrm{s}$ as the time when the blebbing started.
 As a result of the successive blebbing, the oil droplet sometimes exhibited fission.
 After the duration time of 20--70 min depending on STAC concentration, the fission ceased eventually.
 The duration time tended to be longer for the higher STAC concentration.
 
 The snapshots of the droplets for various $C$ are shown in Fig.~\ref{snap}(a).
 Regardless of the STAC concentration, for the first several hundreds of seconds, the oil droplet exhibited the dynamic blebbing, which sometimes induced the droplet fission as shown in the first four columns of the images from the left ($T \leq 360\, \mathrm{s}$) in Fig.~\ref{snap}(a).
 After the period with the droplet fission, the droplet ceased the fission but continued small blebbing as shown in the last columns of the images, on the righ ($T = 7200\, \mathrm{s}$) in Fig.~\ref{snap}(a).
 The fission rate and the final number of the droplets depended on the STAC concentration.
 To quantify the STAC-concentration dependences of the fission dynamics, we analyzed the time series of the number of the oil droplets produced by fission as shown in Fig.~\ref{snap}(b).
 We obtained the ensemble data through ten trials to evaluate the behaviors of the oil droplets and to confirm the reproducibility.
 The mean number of droplets, $\langle N(T)\rangle$, and the individual data are shown with a thick line and thin lines, respectively, in Fig.~\ref{snap}(b).
 
\begin{figure}
    \centering
    \includegraphics{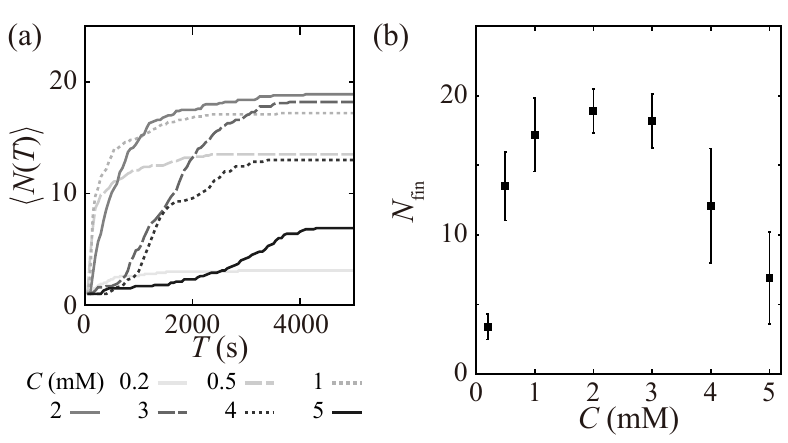}
    \caption{The number of droplets for various $C$. (a)~Time series of the average number of oil droplets, $\langle N(T)\rangle$, for various $C$. (b)~The final number of oil droplets, $N_{\mathrm{fin}}$, for various $C$. $N_{\mathrm{fin}}$ has a peak at $C= 2\, \mathrm{mM}$. Error bars represent standard deviations.}
    \label{Nf}
\end{figure}

 Figure~\ref{Nf}(a) shows $\langle N(T)\rangle$ for various $C$.
 For $C=0.2$--$2\,\mathrm{mM}$, $\langle N(T)\rangle$ rapidly increased during the first $1000$ s, and then converged to each final value depending on $C$.
 On the other hand, for $C=3$--$5\,\mathrm{mM}$, $\langle N(T)\rangle$ slowly increased during the first $1000$ s, then started rapid increases, and finally converged to each final value.
 Figure~\ref{Nf}(b) shows the final number of the droplets $N_\mathrm{fin}$ versus $C$.
 For the lower concentrations $C=0.2$--$2\,\mathrm{mM}$, $N_\mathrm{fin}$ increased with an increase in $C$.
 For the higher concentrations $C=2$--$5\,\mathrm{mM}$, in contrast, $N_\mathrm{fin}$ decreased with an increase in $C$.
 In other words, $N_\mathrm{fin}$ had a peak at $C=2\,\mathrm{mM}$.

\section{Discussion}
\label{discussion}

 In our experimental results, $N_\mathrm{fin}$ has a peak at $C=2\,\mathrm{mM}$ as shown in Fig.~\ref{Nf}(b).
 To discuss the underlying mechanism for the concentration dependence, we investigated the deformation of the oil droplets because our experimental observation suggests the correlation between deformation and fission.
 
\begin{figure}
    \centering
    \includegraphics{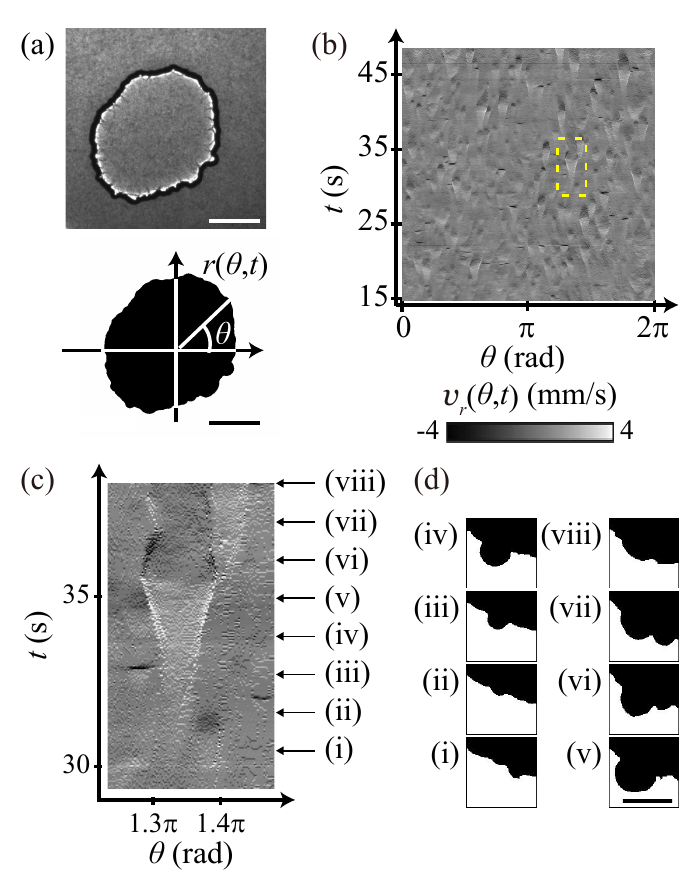}
    \caption{Dynamics of the droplet deformation for $C= 4\, \mathrm{mM}$. (a)~Binarization of the droplet image and definition of the polar coordinates. Scale bar: $10\, \mathrm{mm}$. The center of mass of the droplet was set to be the origin. (b)~Spatiotemporal map of the radial component of the boundary velocity $v_{r}(\theta,t)$. (c)~Magnification of the region in the yellow dashed rectangle in (b). (d)~Snapshots of the binarized images of a single blebbing dynamics corresponding to (c) every $1\,\mathrm{s}$. Scale bar: $5\,\mathrm{mm}$.}
    \label{po_co}
\end{figure}

 Here, we analyzed the concentration-dependent deformation observed in experiments equivalent to those in Sec.~\ref{result} except for the time resolution.
 The deformation was captured at 30 fps.
 Here we defined $t=0\,\mathrm{s}$ as the time when we put the oil droplet on the aqueous surface, and we analyzed the droplet deformation from $t=15\,\mathrm{s}$.
 The reason why we carried out the analyses of the droplet shape just after the induction period instead of just before the fission is described in Appendix~\ref{appendixA}.
 The captured images were processed as follows.
 First, we binarized the images to extract the droplet region, and defined the polar coordinates $(r,\theta)$, where the origin was at the center of mass of the droplet as shown in Fig.~\ref{po_co}(a).
 In these coordinates, we defined the boundary position $r(\theta,t)$ as the largest distance in $\theta$ direction from the origin at $\theta = \theta_n$ and $t=t_m$(see Appendix~\ref{appendixA}).
 Here, $\theta_n = 2\pi n / N_\theta \, \mathrm{rad} \, (n=0,1,\cdots,N_\theta -1)$ and $t_m = m / 30\,\mathrm{s}\,(m=0,1,\cdots,N_t -1)$, where $N_\theta = 1024$ and $N_t = 1024$.
 The droplet just after the induction period is close to circular, and the description of the boundary position by $r(\theta,t)$ is valid (see Appendix~\ref{appendixA} for the details).
 We calculated the radial component of the boundary velocity $v_{r}(\theta,t)$ also at $\theta = \theta_n$ and $t = t_m$ as $v_r(\theta_n,t_m)=[r(\theta_n,t_{m+1})-r(\theta_n,t_m)]/(t_{m+1}-t_m)$, and  plotted the spatiotemporal map of $v_{r}(\theta,t)$ in the same manner as the previous study\,\cite{soft11} in Fig.~\ref{po_co}(b).
 Figure~\ref{po_co}(c) is the magnification of the region in the yellow dashed rectangle in Fig.~\ref{po_co}(b), which corresponds to a single cycle of the expansion and shrinkage, i.e., a single blebbing dynamics.
 As a typical blebbing dynamics, the arc length of the boundary with a positive outward velocity increases, and a bright triangular region appears in the spatiotemporal map.
 The detailed explanations of the correspondence between the blebbing dynamics and the two-dimensional patterns in the spatiotemporal map are described in Appendix~\ref{appendixB}.
 Note that the brighter triangular shape suggests that the edges of the blebbing region are not pinned but they expand with time as shown in Fig.~\ref{po_co}(d).
 In other words, the bleb expands almost in a circular manner.
 After the expansion, the expanded part shrinks rapidly and a dark region smaller than the brighter triangular region appears above it in the map.
 Meanwhile, the maximum lengths of the triangular regions along a horizontal axis correspond to the maximum bleb size measured with the angle.
 The maximum lengths of bright and dark regions along a vertical axis represent the timescale of expansion and shrinkage, respectively.
 Over the entire droplet, a lot of similar patterns are incoherently distributed as seen in Fig.~\ref{po_co}(b), which is consistent with the previous report\,\cite{soft11}.
 Based on the analyses, we distinguished the blebbing and the global deformation.
 Here, the global deformation is defined as the deformation with the larger spatial scale and longer temporal scale compared with the bleb, and it is not related to the pair of the expansion and shrinkage.
 In the followings, we investigated the spatial and temporal scales of the blebbing.
 
\begin{figure}
    \centering
    \includegraphics{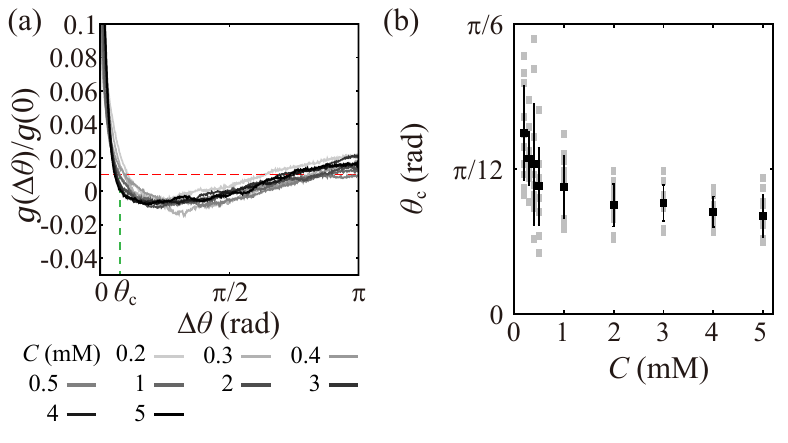}
    \caption{(a)~Autocorrelation function, $g(\Delta\theta)$, of the boundary velocity $v_{r}(\theta, t)$ for various $C$. See Sec.~\ref{discussion} for the detailed definition of $g(\Delta\theta)$. Here, the correlation angle $\theta_{\mathrm{c}}$ was defined as the smallest $\Delta\theta$ that satisfies $g(\Delta\theta)/g(0)=0.01$. In this graph, each line is plotted from the average of ten sets of experimental data. (b)~$\theta_{\mathrm{c}}$ versus $C$. Gray dots are individual data. Black dots with error bars are the averaged values over the ten trials. Error bars represent standard deviations.}
    \label{an_c}
\end{figure}

 We calculated the autocorrelation function $g(\Delta \theta)=\langle v_{r}(\theta,t)v_{r}(\theta +\Delta \theta,t)\rangle_{\theta,t}$ as shown in Fig.~\ref{an_c}(a), where $\Delta \theta$ represents angle difference and $\langle \cdot \rangle_{\theta,t}$ represents the average over $\theta$ and $t$.
 The correlation angle $\theta_\mathrm{c}$ was defined as the smallest $\Delta \theta$ that satisfied $g(\Delta \theta)/g(0)=0.01$.
 Figure~\ref{an_c}(b) shows the $C$ dependence of the correlation angle $\theta_\mathrm{c}$.
 $\theta_\mathrm{c}$ decreased with an increase in $C$.
 
\begin{figure}
    \centering
    \includegraphics{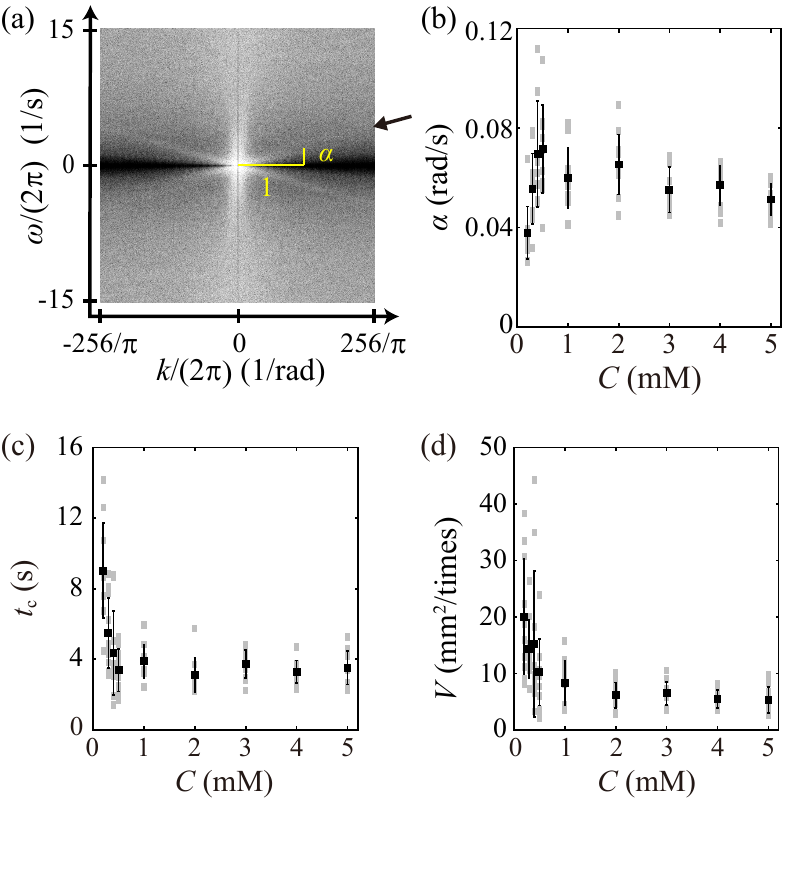}
    \caption{Analyses of the blebbing using FFT images. (a)~FFT image of Fig.~\ref{po_co}(b). Two white inclined lines crossing at the origin correspond to the boundary expansion of the droplet. One of these lines is indicated by the arrow on the right side. We measured the slope $\alpha$ of the white lines. The almost vertical line corresponds to the rapid shrinkage. The contrast of the image was enhanced for visibility. Brighter and darker regions correspond to the higher and lower values of the power in the Fourier spectra. (b)~Expansion rate $\alpha$ versus $C$. (c)~Characteristic timescale of expansion $t_{\mathrm{c}}$ versus $C$. (d)~Bleb size $V$ versus $C$. Error bars represent standard deviations.}
    \label{tc}
\end{figure}

 Next, we estimated the typical timescale of blebbing.
 Only when the droplet shape changes into a distorted shape can the droplet exhibit fission.
 For such global deformation, expansion is necessary rather than shrinkage.
 Thus, we consider that the droplet-boundary expansion would provide the essential information for the droplet fission.
 The timescale of expansion was longer than that of shrinkage as we can see the longer vertical length of the bright region than that of the dark region [Fig.~\ref{po_co}(c)].
 We tried to extract the two timescales from the time-domain autocorrelation function, but it was difficult to distinguish them (see Appendix~\ref{appendixC}.
 Instead, to evaluate the characteristic timescale of expansion $t_\mathrm{c}$, we applied two-dimensional fast Fourier transform (FFT) to $v_{r}(\theta,t)$.
 Figure~\ref{tc}(a) shows the typical result of the two-dimensional FFT, in which the horizontal axis corresponds to the angular wavenumber $k$, and the vertical axis to the angular frequency $\omega$.
 From the characteristic triangular patterns seen in $v_r(\theta,t)$ map [Fig.~\ref{po_co}(c)], we can see that the droplet boundaries expand at a constant rate.
 In the FFT image, the bright lines orthogonal to the original triangular patterns should appear; We can see two white inclined lines and a white vertical band in Fig.~\ref{tc}(a).
 The inclined lines with the slope of $\alpha$ correspond to the triangular bright regions in Fig.~\ref{po_co}(c).
 Notably, $\alpha$ corresponds to the expansion rate of the central angle of the bleb (see Appendix~\ref{appendixB}).
 The white vertical band corresponds to the dark region on the bright triangular region in the original spatiotemporal map.
 The detailed explanations for extracting the correlation time from the FFT image are described in Appendix~\ref{appendixD}.
 
\begin{figure}[tb]
    \centering
    \includegraphics{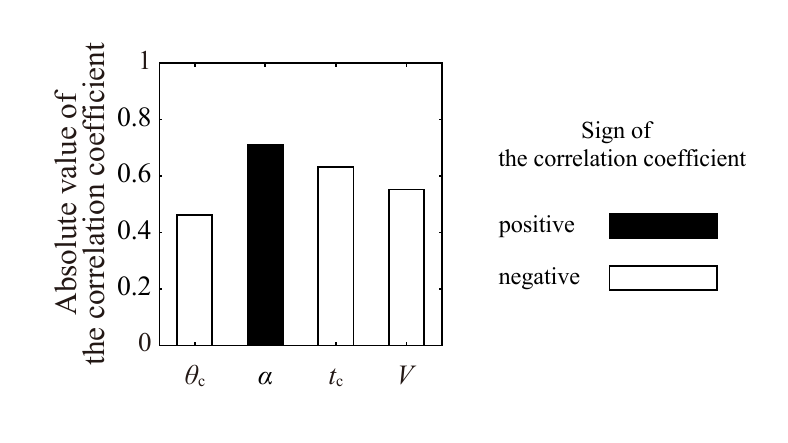}
    \caption{Absolute values of the correlation coefficients between the final number of the oil droplets $N_{\mathrm{fin}}$ and one of the four parameters: correlation angle $\theta_{\mathrm{c}}$, characteristic timescale of expansion $t_{\mathrm{c}}$, expansion rate of the central angle of the bleb $\alpha$, and the bleb size $V$. Filled and open bars represent positive and negative correlations, respectively.}
    \label{cv}
\end{figure}

 Figure~\ref{tc}(b) shows the $C$ dependence of the expansion rate $\alpha$.
 For $C<2\,\mathrm{mM}$, $\alpha$ increased with an increase in $C$.
 For $C>2\,\mathrm{mM}$, $\alpha$ remained almost constant at $\alpha \approx 0.06$~rad/s.
 We calculated the characteristic timescale of expansion $t_\mathrm{c}$ as $t_\mathrm{c}=\theta_\mathrm{c} / \alpha$.
 Figure~\ref{tc}(c) shows the $C$ dependence of $t_\mathrm{c}$.
 Qualitatively similar to the $C$ dependence of $\theta_\mathrm{c}$, $t_\mathrm{c}$ decreased with an increase in $C$.
 The bleb size $V$ was calculated as $V=R^2\theta_\mathrm{c}t_\mathrm{c}\alpha$, and its $C$ dependence is shown in Fig.~\ref{tc}(d).
 To obtain the bleb size $V$, we assumed the concentric-circle-like expansion.
 The average radius of the droplet $R$ was estimated by $R=\sqrt{S/\pi}$, where $S$ is the droplet area, and we found it almost constant at $R\approx 12\,\mathrm{mm}$ from the experiments.
 $V$ also decreased with an increase in $C$.
 Then, we calculated the correlation coefficients between $N_{\mathrm{fin}}$ and one of the parameters $\theta_{\mathrm{c}}$, $\alpha$, $t_{\mathrm{c}}$, and $V$ from their $C$ dependence, as shown in Fig.~\ref{cv}.
 We calculated the correlation coefficient for given data set $\{x\}=\{x_1, x_2,\ldots,x_n\}$ and $\{y\}=\{y_1, y_2, \ldots,y_n\}$ as $S_{xy}/(S_{x}S_{y})$, where the covariances of $x$ and $y$ are defined as $S_{xy}=\Sigma^{n}_{i=1}(x_{i}-\overline{x})(y_{i}-\overline{y})/n$, and standard deviations of $x$ and $y$ are defined as $S_{x}=\sqrt{\Sigma^{n}_{i=1}(x_{i}-\overline{x})^2 /n}$ and  $S_{y}=\sqrt{\Sigma^{n}_{i=1}(y_{i}-\overline{y})^2 /n}$, respectively.
 Here, the overline denotes the average of the data and $n$ is the number of the data of $x$ and $y$.
 In this study, $x$ corresponds to $N_\mathrm{fin}$, and $y$ corresponds to $\theta_\mathrm{c}$, $\alpha$, $t_\mathrm{c}$, or $V$.
 To obtain the dimensionless value, we used the correlation coefficient $S_{xy}/(S_{x}S_{y})$.
 
 From these results, we confirmed that the correlation coefficient between $N_{\mathrm{fin}}$ and $\alpha$ was positive and the absolute value of it was the largest among the four.
 The correlation coefficients between $N_{\mathrm{fin}}$ and the other parameters were negative and the absolute values were smaller.
 Through the correlations between $N_{\mathrm{fin}}$ and these parameters, it seems that the droplet exhibits fission more frequently when the droplet exhibits faster expansion with a smaller bleb size, and when the timescale of expansion is shorter.
 To test the null hypothesis of no correlations, we also calculated the $p$ values of the four parameters by using Student's $t$-test\,\cite{cgac68}.
 The $p$ values for $\theta_{\mathrm{c}}$, $\alpha$, $t_{\mathrm{c}}$, and $V$ were $0.295$, $0.074$, $0.128$, and $0.199$, respectively.
 The $p$ value of the correlation coefficient between $N_{\mathrm{fin}}$ and $\alpha$ was the smallest among the four, and thus it had the strongest correlation with $N_{\mathrm{fin}}$ among the four.
 
 For lower $C$ ($C<0.4\,\mathrm{mM}$), the bleb size $\theta_{\mathrm{c}}$ decreased with the increase in $C$ but the expansion rate $\alpha$ increased.
 For higher $C$ ($C>2\,\mathrm{mM}$), $\theta_{\mathrm{c}}$ and $\alpha$ slightly decreased with $C$.
 From these experimental results, we considered the fission mechanism depending on $C$ as follows.
 For lower $C$, the gel layer formed at the interface should be thinner\,\cite{jpcb09}.
 The thinner gel layer not only generates smaller driving force but also has smaller rigidity, resulting in the lower $\alpha$ and larger $\theta_{\mathrm{c}}$.
 For higher $C$, the gel layer should be thicker\,\cite{jpcb09}, and the thicker gel layer may generate larger driving force and stiffen the interface.
 The increase in the rigidity of the thicker gel layer might become dominant compared to the increase in the driving force, resulting in the lower $\alpha$ and $\theta_{\mathrm{c}}$.
 The expansion rate $\alpha$, which is determined by the driving force and the material viscoelasticity, reflects the competition between the driving force and the deformability, and thus may have the strongest contribution to the deformation and fission.
 Actually, despite the remarkable negative correlation obtained between $N_{\mathrm{fin}}$ and $\theta_{\mathrm{c}}$, between $N_{\mathrm{fin}}$ and $\theta_{\mathrm{c}}$-dependent values ($t_{\mathrm{c}}$ and $V$) as shown in Fig.~\ref{cv}, the absolute values of the correlation coefficients for these three parameters were relatively smaller than that for $\alpha$.
 Further studies for the effect of such competitive effects between the driving force and the deformability is needed to clarify more detailed mechanisms underlying the droplet deformation and fission.
 
\section{Conclusion}
\label{conclusion}
 We studied the fission of the tetradecane droplet containing PA on the STAC aqueous solution for various STAC concentrations.
 The number of the oil droplets at the final stage of the observation had a peak at a certain concentration.
 To discuss the droplet fission dynamics, we extracted the boundary velocity from the time series of the droplet deformation and calculated the correlation in time and space for various STAC concentrations.
 We obtained the four parameters that characterize the blebbing: the correlation angle, the characteristic timescale of expansion, the expansion rate, and the bleb size.
 Among them, the expansion rate had the strongest positive correlation with the final number of the droplets and the others had negative correlations.
 Therefore it is suggested that faster, smaller-size, and shorter-time expansion leads more frequent fission.
 It is known that the generation and rupture of the active gel are also seen in the migration, deformation and fission of living cells\,\cite{cell96,tcb03,nat05,naph15}.
 The present results might also help understand such phenomena by considering the common features.
 
\begin{acknowledgments}
This work was supported by JSPS KAKENHI Grant Nos. JP19H00749, JP19H05403, JP19K14675, JP16H03949. It was also supported by Sumitomo Foundation (No.~181161), the Japan-Poland Research Cooperative Program between JPSJ ans PAN Grant No. JPJSBP120204602, and the Cooperative Research Program of ``Network Joint Research Center for Materials and Devices'' (No.~20191030).
\end{acknowledgments}

\appendix
 \section{Elapsed-time dependence of the droplet deformation}
 \label{appendixA}
 We considered that the droplet exhibited the deformation larger than the typical bleb size through continuous local blebbings, and such global deformation results in the droplet fission.
 Thus, we analyzed the blebbing dynamics to investigate the droplet fission related to the blebbing dynamics.
 $t=0\,\mathrm{s}$ was set as the time when the droplet was put on the aqueous surface.
 We focused on the blebbing dynamics after $t=15\,\mathrm{s}$ (earlier stage) rather than the blebbing dynamics just before the fission (later stage).
 This section gives more detailed explanations why we carried out the analyses of the blebbing dynamics at the earlier stage.
 
\begin{figure*}[bt]
     \centering
     \includegraphics[width=17cm]{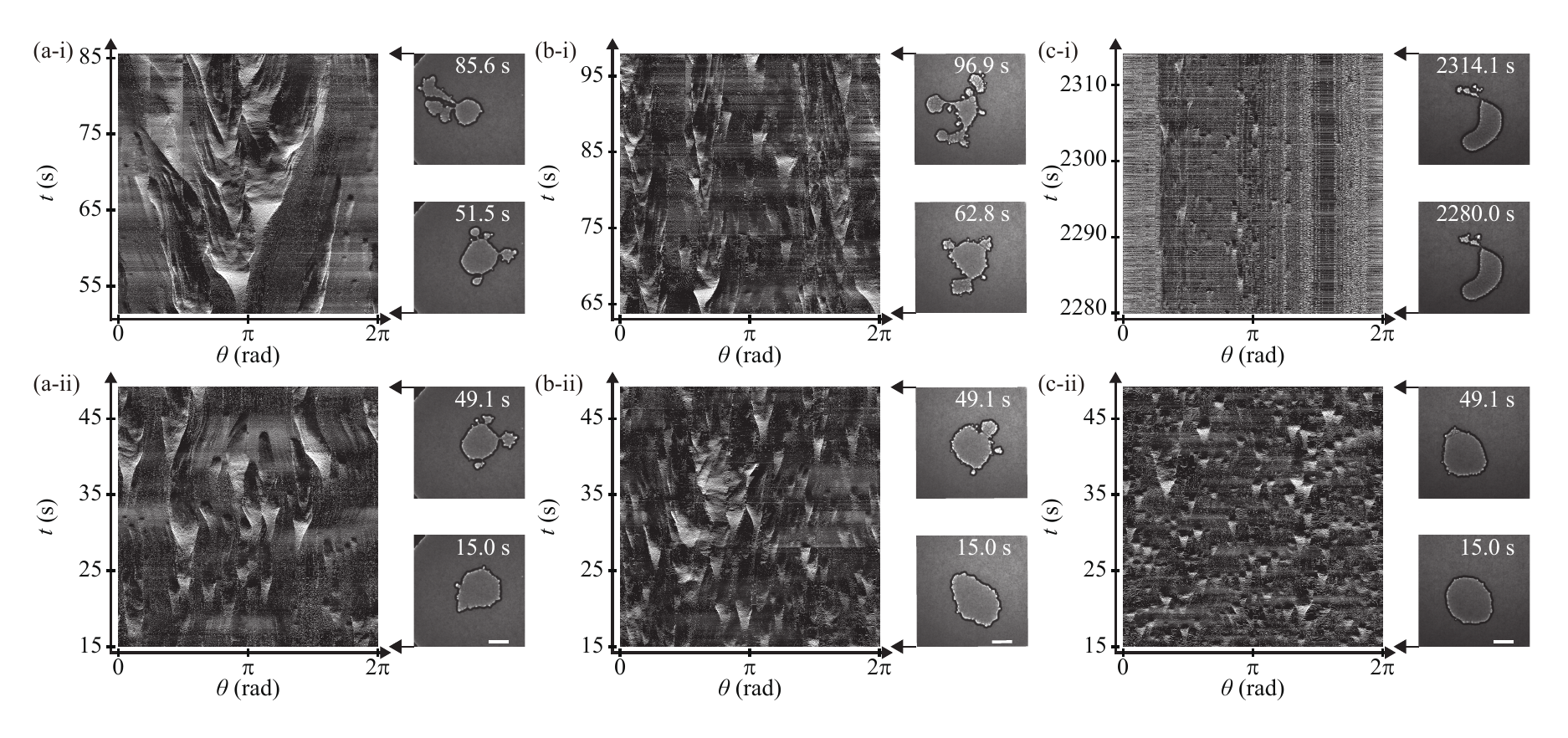}
     \caption{Spatiotemporal maps and the snapshots. Spatiotemporal maps and the snapshots of the droplet just before fission (later stage) at $C=$ (a-i)~$0.5\,\mathrm{mM}$, (b-i)~$2\,\mathrm{mM}$, and (c-i)~$5\,\mathrm{mM}$, respectively. $t=0\,\mathrm{s}$ was defined as the time when the droplet was put on the aqueous surface. Spatiotemporal maps and the snapshots of the droplet from $t=15\,\mathrm{s}$ (earlier stage) at $C=$ (a-ii)~$0.5\,\mathrm{mM}$, (b-ii)~$2\,\mathrm{mM}$, and (c-ii)~$5\,\mathrm{mM}$, respectively. Similar bright triangular and dark small regions are observed in both the stages; for example, the regions observed in the maps (a-i) and (a-ii) as well as the snapshots for 15.0 s and 85.6 s, indicating the similar-scaled blebbing in both the stages. Scale bars: 10 mm.}
     \label{fis}
 \end{figure*}

 Figure~\ref{fis} shows the spatiotemporal maps and the snapshots of the droplet in both the earlier stage and the later stage for various STAC concentrations $C$.
 The droplet shape at the earlier stage was circular [see $15\,\mathrm{s}$ in Figs.~\ref{fis}(a--c)], but the one at the later stage was distorted [see $85.6\,\mathrm{s}$, $96.9\,\mathrm{s}$, and $2314.1\,\mathrm{s}$ in Figs.~\ref{fis}(a), \ref{fis}(b), and \ref{fis}(c), respectively].
 
 As mentioned in Sec.~\ref{discussion}, we set the polar coordinates and the origin was set at the center of mass of the droplet in order to define the boundary position of the droplet.
 However, in the distorted droplet, we detected multiple droplet boundaries in a certain radial direction as shown in Fig.~\ref{fis}(a-i).
 Thus, we defined $r(\theta,t)$ as the largest distance from the origin to the droplet boundary in $\theta$ direction, and its time derivative, i.e., the radial component of the boundary velocity, $v_r(\theta,t)$, was calculated to obtain the spatiotemporal map.
 The proper droplet shape should be reflected only when the droplet shape is close to circular, since the radial direction does not always meet the normal direction of the droplet boundary for the intensively distorted droplets in the later stage as shown in Figs.~\ref{fis}(a-i), (b-i), and (c-i).
 In addition, even for the same-sized bleb with the same peripheral length $r(\theta,t)\theta_\mathrm{c}$, $\theta_\mathrm{c}$ in a certain direction can be different from that in other directions in a distorted droplet, since $r(\theta,t)$ intensively depends on $\theta$.
 From these reasons, $r(\theta,t)$ at the earlier stage correctly reflected the blebbing dynamics but the one at the later stage could not, and thus the one at the later stage was not suitable for the analyses.
 
 From the snapshots of the droplets in Figs.~\ref{fis}(a--c), we confirmed that the bleb size at the earlier stage was almost the same as the one at the later stage.
 Thus, to obtain the information of the deformation, we analyzed the blebbing dynamics at the earlier stage.

 \section{Correspondence between blebbing and Spatiotemporal map}
 \label{appendixB}
 A typical cycle of blebbing, i.e., expansion and subsequent shrinkage of a local boundary of a droplet, is identified as the combination of bright triangular and dark small regions in the spatiotemporal map as shown in Figs.~\ref{po_co}(b) and \ref{po_co}(c).
 Here, we explain the relationship between the characteristic patterns in the map and the blebbing dynamics.
 
\begin{figure}
     \centering
     \includegraphics{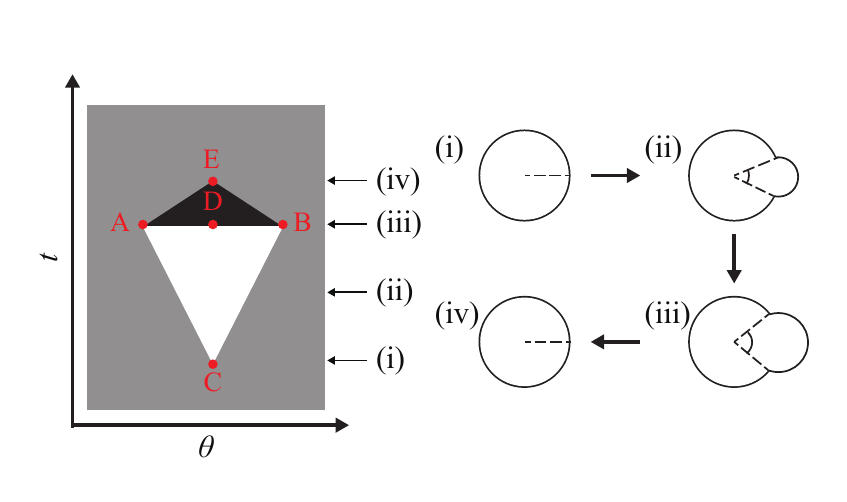}
     \caption{Schematic illustrations of a typical cycle of the blebbing represented in the spatiotemporal map and the corresponding droplet shapes. The typical angular size of the bleb 
     and the typical timescale of the boundary expansion correspond to the lengths of the line segments AB and CD, respectively. The expansion rate $\alpha$ of the central angle of the bleb corresponds to the angle formed by the line segments CB and CD.}
     \label{stmap}
 \end{figure}

 Figure~\ref{stmap} shows the schematic illustration on a typical cycle of the blebbing represented in the spatiotemporal map and the corresponding process of the blebbing.
 In the spatiotemporal map, the bright and dark regions represent the outward expansion with a positive radial velocity $v_{r}(\theta,t)>0$ and the inward shrinkage with $v_{r}(\theta,t)<0$, respectively. 
 The other static parts with $v_{r}(\theta,t)\simeq 0$ appear as the gray regions in the map.
 The initially circular droplet (i) starts to deform by the increase in the inner pressure, and (ii) forms a locally expanded part, called a bleb, until the outward expansion (iii) leads to the rupture of the gel layer at the interface.
 The angle measured from the droplet center of mass, which is indicated by the dashed lines, increases during the expansion from (i) to (iii).
 This angle is referred to as the central angle of the bleb.
 Since the outward expansion is associated with the positive radial velocity $v_{r}(\theta,t)$, the bright triangular region as bounded by the vertices ABC appears in the spatiotemporal map in Fig.~\ref{stmap}.
 The relevant spatiotemporal scales are the typical angular size of the bleb and the timescale of the expansion, which correspond to the line segments AB along the (horizontal) $\theta$ axis and CD along the (vertical) $t$ axis in the map, respectively.
 The angle between the line segments CD and CB in the map corresponds to the expansion rate $\alpha$ of the central angle of the bleb.
 After the rupture of the gel layer, the boundary of the droplet (iv) rapidly shrinks due to the interfacial tension at the oil-water interface, and the  central angle of the bleb decreases with time.
 The negative radial velocity $v_{r}(\theta,t)$ during the inward shrinkage results in the dark region as indicated by the vertices ABE in the spatiotemporal map in Fig.~\ref{stmap}.
 According to these correspondences, the blebbing dynamics appears as a pair of the bright triangular region and the neighboring dark small region in the spatiotemporal map.
 
 \section{Time-domain autocorrelation function}
\label{appendixC}
\begin{figure}
    \centering
    \includegraphics{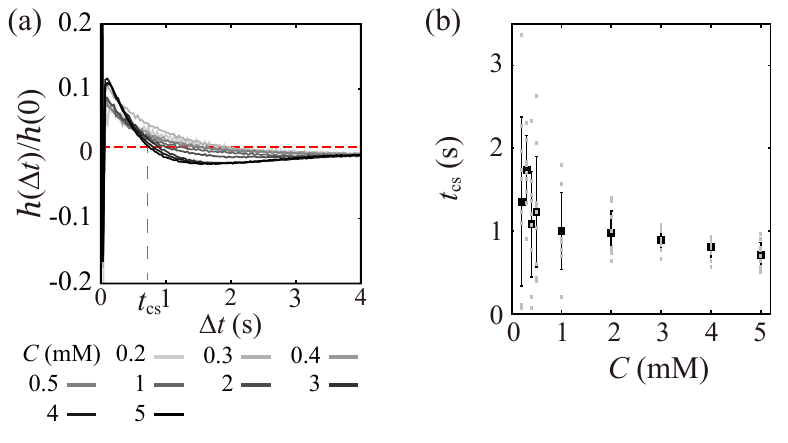}
    \caption{(a)~Time-domain autocorrelation function, $h(\Delta t)$, of the boundary velocity $v_{r}(\theta,t)$ for various $C$. Detailed definition of $h(\Delta t)$ is in the text. The third smallest $\Delta t$ that satisfied $h(\Delta t)/h(0)=0.01$ is set as $t_{\mathrm{cs}}$. In this graph, each line is plotted from the average of ten sets of experimental data. (b)~$t_\mathrm{cs}$ versus $C$. Error bars represent standard deviations. Black dots with error bars are the averaged values over the ten trials.}
    \label{shrink}
 \end{figure}

 We obtained the time-domain autocorrelation function $h(\Delta t)=\langle v_{r}(\theta,t)v_{r}(\theta,t+\Delta t)\rangle_{\theta,t}$ shown in Fig.~\ref{shrink}(a).
 $h(\Delta t)$ for each $C$ became negative where $\Delta t=1/30\,\mathrm{s}$, the minimum time interval, because $h(\Delta t)$ is calculated from the boundary velocity calculated from the time difference of the boundary position with pixel noises.
 That is to say, the velocities at two time points with the minimum time interval tend to have the negative correlation by such pixel noises.
 Therefore we detected the third smallest $\Delta t$ that satisfied $h(\Delta t)/h(0)=0.01$.
 This $\Delta t$, represented as $t_{\mathrm{cs}}$, is a candidate for the characteristic timescale of the expansion, which is plotted in Fig.~\ref{shrink}(b) for each $C$.
 The typical order of $t_{\mathrm{cs}}$ is the same as $t_{\mathrm{c}}$ shown in Fig.~\ref{tc}(c).
 However, there should be two significant timescales in the blebbing deformation: those of expansion and shrinkage.
 It was difficult to judge whether $t_{\mathrm{cs}}$ corresponds to the timescale of expansion, as we could not find any remarkable feature for the two different timescales greater than $1/30\,\mathrm{s}$ in Fig.~\ref{shrink}(a).
 This is the reason why we did not use $h(\Delta t)$ to analyze the timescale of expansion.


 \section{FFT image of spatiotemporal map}
 \label{appendixD}
  \begin{figure}[h]
    \centering
    \includegraphics{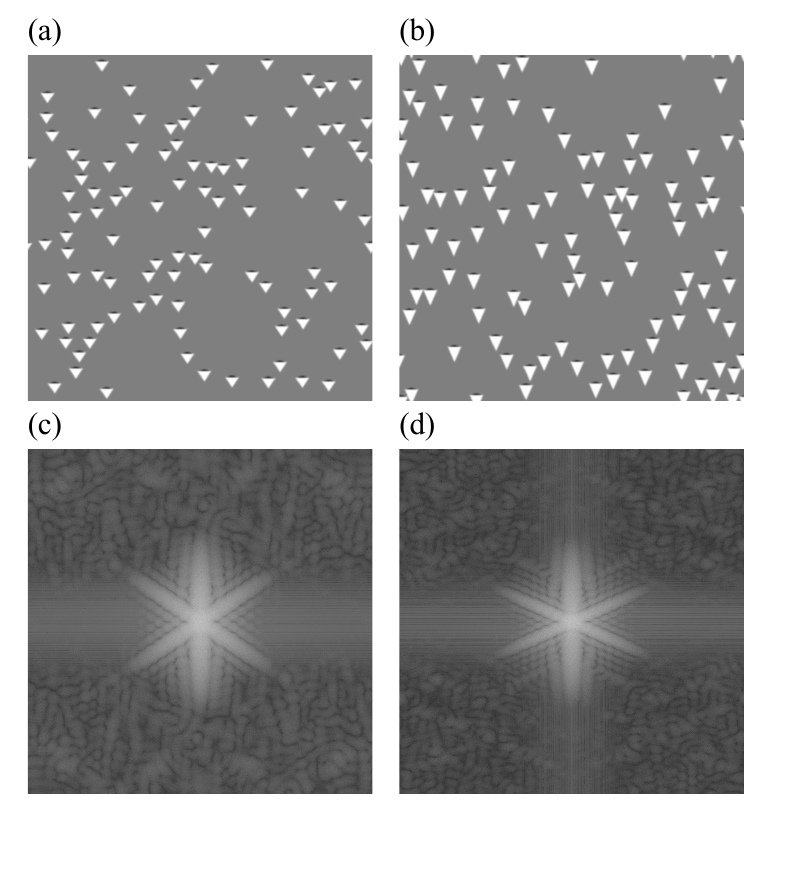}
    \caption{Demonstrative FFT analyses with artificially generated images. (a,b)~Artificially generated spatiotemporal maps with different lengths of the white regions along the vertical axis, which correspond to the analysis of the blebbing dynamics in Fig.~\ref{po_co}(b). Detailed setup of these maps are written in the text. (c,d)~FFT images of (a,b), respectively.}
    \label{fftpre}
\end{figure}
 We demonstrate the correspondence of the two-dimensional patterns in the spatiotemporal map and those in the FFT image shown in Figs.~\ref{po_co}(b) and \ref{tc}(a).
 Figures~\ref{fftpre}(a) and \ref{fftpre}(b) are the artificially generated spatiotemporal maps that correspond to the random deformation of the droplet boundary without noise.
 The lengths of white regions along the horizontal axis in Fig.~\ref{fftpre} are $40$ pixels, and those along the vertical axis in Figs.~\ref{fftpre}(a) and \ref{fftpre}(b) are $35$ and $45\,\mathrm{pixels}$, respectively.
 The difference in the length along a vertical axis corresponds to that in the timescale of expansion.
 The gray region represents stationary boundary, the white ones represent boundary expansion, and the black ones represent boundary shrinkage.
 A pair of white and black regions corresponds to a single blebbing and is called a blebbing area.
 In these maps, 100 blebbing areas are scattered in a spatiotemporally random manner.
 Gaussian blur with the standard deviation of two pixels is applied to the both of $1024\times1024$ images.
 Figures~\ref{fftpre}(c) and \ref{fftpre}(d) correspond to the FFT images of Figs.~\ref{fftpre}(a) and \ref{fftpre}(b), respectively.
 Note that the FFT image corresponds to the ``structure factor'' in the field of crystallography.
 In both of these images, a vertical white line and two inclined white lines can be seen.
 As indicated in Figs.~\ref{fftpre}(c) and \ref{fftpre}(d), the different lengths of the white region along the vertical axis result in the different slopes of the two inclined lines, referred to as $\alpha$.
 Thus, the two inclined lines correspond to the structure factor of the white regions.
 
 In the FFT images, we can also find a white vertical band as seen in the experimental data in Fig.~\ref{tc}(a).
 In the analyses of the experimental data, we subtracted the translational motion of the droplet.
 Consequently, the first spatial mode ($\pm 1$ mode) of the FFT images is absent, which is represented as the two vertical black thin lines in Fig.~\ref{tc}(a).
 Because the artificially generated spatiotemporal maps in Figs.~\ref{fftpre}(a,b) include the first mode, such two vertical black lines are absent in Figs.~\ref{fftpre}(c,d).

\end{document}